# Managing Complexity

David P. CHASSIN
Energy Science and Technology Division, Pacific Northwest National Laboratory
Richland, Washington 99352

and

Christian POSSE and Joel MALARD
Fundamental Sciences Division, Pacific Northwest National Laboratory
Richland, Washington 99352

## ABSTRACT

Physical analogs have shown considerable promise for understanding the behavior of complex adaptive systems, including macroeconomics, biological systems, social networks, and electric power markets. Many of today's most challenging technical and policy questions can be reduced to a distributed economic control problem. Indeed, economically-based control of large-scale systems is founded on the conjecture that the price-based regulation (e.g., auctions, markets) results in an optimal allocation of resources and emergent optimal system control. This paper explores the state of the art in the use physical analogs for understanding the behavior of some econophysical systems and deriving stable and robust control strategies for them. In particular we review and discuss applications of some analytic methods based on a thermodynamic metaphor according to which the interplay between system entropy and conservation laws gives rise to intuitive and governing global properties of complex systems that cannot be otherwise understood.

## INTRODUCTION

The idea of controlling complex systems using decentralized econophysical processes is not new, nor has it always been done with a literal market in the sense that product price and quantity are the joint means of determining the allocation of a scarce resource. Indeed, in the early 1980s, so-called contract networks were demonstrated to allocate scarce central processing unit (CPU) time to competing tasks in high-performance computers [1]. Nevertheless, the challenges faced by system designers who wish to use economically-efficient control strategies are daunting. Such systems are generally called complex adaptive systems because they exhibit a property that is regarded as a profound obstacle to cost-effectively crafting stable and robust classical control strategies: emergent behavior [2] [3] [4] [5]. This is the property of systems that causes them to exhibit global behaviors not anticipated by the control strategy. For example, building heating, ventilation, and air-conditioning (HVAC) systems exhibit a phenomenon called global hunting, an accidental artifact of the control strategies employed to govern the thermodynamic process [6]. Global hunting exhibits itself as an unexpectedly prolonged cycling between two or more local minima.

Many systems, in economics [7] [8] [9], environmental sciences [10], molecular biology [11], power engineering [12], sociology [13], military command/control/communications/intelligence ($C^3I$) [14], and transportation systems, exhibit such unanticipated emergent behavior and are thus candidates for membership in the class of systems characterized by complex adaptive behavior with the potential for unanticipated robustness degradation [15] [16]. However, exploiting the emergent behavior of complex systems [17] by design has been the subject of discussions with respect to building controls [18] and using price as the only signal for electric power systems control [19]. The challenge to exploiting such phenomena is predicting the general characteristics of emergent behavior in complex systems as they related to the behavior of individual components. This is fraught with analytic difficulties, many of which are discussed by Atmanspacher and Wiedenmann [20].

It is with the object of understanding the relationship between the rules governing the behavior of system components and the emergent econophysical behaviors of the whole systems that we seek a theory of control based on the thermodynamic analogy [21]. Tsallis et al. [22] outlined how one might apply non-extensive statistical mechanics to the question of stability and robustness. Donangelo and Sneppen [23] went further to discuss the dynamics of value exchange in complex systems. In economics, Smith and Foley [7] show that the correspondence of utility theory to thermodynamics defines a whole consistent methodology, and not just a set of analogies. Sergeev [8] and Jaynes [9] address the central role of entropy in a thermodynamics-inspired approach to market equilibrium, and how it should be used. Jaynes [24] further expands the analogy for time-varying phenomena using the formalism of predictive statistical mechanics. But Fleay [25] argues that the conditions imposed by a simplistic thermodynamic approach are utopian and can never be met for real power markets. Nonetheless, the first step in any attempt to understand large-scale system behavior and control is to devise a rigorous, albeit pedagogic, time-independent model of aggregate properties of a system based on those of individual components. *A priori*, it is by no means certain that the analogy to statistical mechanics should hold strictly, nor is it necessary. But insofar as it does, we can proceed with the derivation of such global properties as are consistent with it. Pragmatically we can conclude that the analogy is true if we find no inconsistencies with empirical data or existing theories and if the aggregate properties we develop usefully elucidate the behavior of the systems being considered.

## METAPHORS FOR COMPLEX SYSTEMS

Physics metaphors have been repeatedly used to model and understand the conceptual and methodological problems arising in other sciences. Classical economics was built largely on the analogy to mechanics. Adam Smith's "invisible hand" is a vivid metaphor combining market coordination of selfish individuals scattered throughout a country or even around the world with the idea of an anonymous will pushing towards the social good. This parallels the focus of modern analysis on equilibrium, decentralization and efficiency. This was formalized a century later by Leon Walras in providing the first mathematical statement of general equilibrium theory with rational mechanics.

When Adam Smith wrote The Wealth of Nations, the principles of thermodynamics were not known yet, so the "invisible hand" was interpreted in terms of the era, i.e., a mechanical metaphor. Adam Smith's conceptual model, which identifies the effect of human motivation with "forces" on the market, gives serious grounds for such an interpretation. In the mechanical models, equilibrium is that state at which the forces applied to the system counterbalance each other and the potential, when steady, reaches its extremum. Consequently, to apply the mechanical metaphor of equilibrium in the economics, some analogs of the mechanical notions are needed. Such a conceptualization is not "harmless" at all; it implies that the system, having slightly digressed from the state of equilibrium, will return to this very state being left alone. But how this occurs is not stated.

Sergeev seems to have set out to address this question when he proposed a more general concept of temperature according to which the equilibrium *temperature* (italics are used to indicate an abstract notion as opposed to the physical quantity) of a system is simply the result of the condition that "there are no [net] flows of the conserved macro-parameters between the parts of the system." Thus he defines the notion of *temperature* as "the derivative of the entropy of the Large System with respect to the macro-parameter for which there exists a conservation law." He makes a particular point to observe that

> «…we said nothing related to either physics, or physical laws and observables. **All the arguments are applicable to Large Systems of any nature, subject to the above hypotheses**. These arguments look rather natural, in relation to the large economic systems, if we regard the total income, the total value of products or the total value of consumption of goods as macro-parameters, whereas distribution of income and products of consumption of goods between the subjects of economic activity are viewed as micro-parameters.» (emphasis added) [8]

Sergeev pursues the analogy further to discuss what he terms *migration potential*, a more economic form of electrochemical potential. He uses it to describe the how agents flow from one economic system to another (as opposed to *temperature*'s influence of product flow). He asserts that together these notions are sufficient to describe the equilibrium dynamics of economic systems, stopping short of developing any model for time-dependent dynamics.

Among the valuable contributions made by Sergeev, Smith and Foley is that according to this method of analysis, one can learn a great deal regarding the general stability of a complex system by considering the form of the system's entropy function. In particular, entropy functions that have multiple maxima are characteristic of systems with bifurcation behavior. In addition, certain higher order entropy functions are characteristic of unstable or runaway systems, whereby in certain circumstances they have no finite maximum entropy. In other circumstances systems can stagnate when the entropy function is flat within a broad range of states.

Time-independent equilibrium models appear relatively easy to develop, so we reasonably expect the physical metaphor to be easily applied to any complex system, biological, military, environmental, engineering or otherwise. However, the challenge raised by Fleay requires that we consider the time-dependent perturbation behavior of these systems. This is a significantly more challenging problem, and it is the ultimate goal of any attempt to model the dynamic behavior of large-scale systems, whether open or closed.

If there were any lingering doubts about the impact of these ideas on economics, we can hope that Focardi and Fabozzi's statement will overcome them

> «More recently, the diffusion of electronic transactions has made available a huge amount of empirical data. The availability of this data created the hope that economics could be given a more solid scientific grounding. A new field—econophysics—opened with the expectation that the proven methods of the physical sciences could be applied with benefit to economics. It was hypothesized that economics systems could be studied as physical systems with only minimal a priori economic assumptions. Classical econometrics is based on a similar approach; but while the scope of classical econometrics is limited to dynamic models of time series, econophysics uses all the tools of statistical physics and complex systems analysis, including the theory of interacting multiagent systems.» [26]

This is as close to an endorsement as is possible today, and it is indeed recognition of the potential impact these ideas have on the science of economics and the broader applications to the design and operation of complex economically Pareto-efficient engineered, biological, and environmental systems.

## THE GRAND CHALLENGE

When adaptable autonomous agents or organisms interact intimately in an environment, such as in predator-prey and parasite-host relationships, they influence each other's evolution. This effect is called co-evolution, and it is the key to understanding how all large-scale complex adaptive systems behave over the long term. Without explicitly pointing it out, Kaufman [2] suggests that there are two distinct scales of co-evolution: inter-species and system-wide. Inter-species co-evolution is the conventional model and ecological examples abound. Numerous examples exist of avian and insect species with co-evolved feeding and breeding strategies that depend on the parasitic or predatory practices of other species. In almost all these cases, it can be shown that advantage is conferred upon both participants, albeit sometimes difficult to discern (as in brood parasites). But, it can be reasonably argued that the presence of another species or even reciprocal changes in response to it are not sufficient evidence for co-evolution and that there must be evidence from quantitative analysis that shows altered or accelerated evolution of individuals deviated from the expected path based on prevailing conditions. Nonetheless, the conditions for co-evolution arising from cooperative-competitive strategies exist in general networks: social networks, *viz.* the use of criminal informants in law-enforcement; economic networks, *viz.* stock trading strategies and the Security and Exchange Commission (SEC) rule-making; electric power, *viz.* by electric power wheeling and Federal Energy Regulatory Commission (FERC) regulatory efforts; military strategy, *viz.* arms races.

System co-evolution is a much more large-scale impact wherein the interaction of one (or a few co-evolved species) with the system as a whole results in changes so fundamental that all species in the system must adapt and the system itself changes in very significant ways. Certainly human interaction with the global environment is a commonly cited example of this phenomenon, although it has yet to shown that our impact is demonstrable in the genetics of most species. But military planners, economists, engineers and biologist also encounter such system-wide adaptive "tipping" phenomena in the systems they study.

The phenomenology of co-evolution in biology is well explored. However, what ultimately governs co-evolution from a first-principles perspective is not entirely clear. The question of whether a single *ab initio* law exists is important to every domain in which complexity is prevalent, whether we consider military strategy, economic policy, global climate change or vaccine development. Certainly entropy maximization and free-energy minimization play important roles in determining which paths a species or a system takes. Morowitz, Bak, Kaufman all proposed a so-called "Fourth Law of Thermodynamics" according to which systems tend to self-organize [2] [27] [28]. The conditions required for this to occur vary depending on the author. However, as in the distinction between Clausius' weak statement that entropy *tends* to increase versus Gibbs' strong statement that entropy *will* increase, we can easily see that any proposed Fourth Law is not a law *per se* until it can be stated unequivocally. This remains to be accomplished. Indeed, it is not clear that a Fourth Law is indeed necessary if it can be easily shown that self-organization and co-evolution are merely a common consequence of the balance required by the competing objectives of maximal entropy and minimal conserved quantities. It has not been shown that these are a necessary cause, let alone a sufficient one. Be that as it may, the strongest definition of the Fourth Law is proposed by Liechti:

> *«If a system receives a through-flow of exergy (produces entropy/dissipate energy), (a) the system will utilize this exergy flow to move away from thermodynamic equilibrium, (b) if [...] more than one pathway to move away is offered from thermodynamic equilibrium, the one yielding most stored exergy, with the most ordered structure and the longest distance to thermodynamic equilibrium, will have a prospensity to be selected.»* [29]

The definition does not match perfectly Kaufman's or Bak's, but it does lead us to consider what happens when multiple systems interact. As a general approach, we should therefore extend our scope to encompass the interaction of system entropy and the minimization of conserved quantities (selected appropriately according to the domain) as the fundamental insight that must be developed. Unfortunately, this is not likely to be sufficient. The main objection is that it does not anticipate the path-dependent behavior that is often observed in self-organizing systems. However, it is clear that Jaynes anticipated this when he proposed a generalization of the Carnot cycle as a simple but more useful model to address the problem of explaining path-dependent behavior in complex systems.

The solution to this problem was identified by Smith and Foley when they proposed using the physical concept of *engine cycles* [7]. In the context of economic theory, they expressed the motivation for this as follows:

> *«Unless the small agents' disequilibrium is constantly replenished, the market maker's activities [...] are more relevant to one-shot arbitrage than to the creation of a sustained pattern of activities. A more interesting question is what can be extracted by a small speculator operating between heterogeneous reservoirs, which cannot trade with each other directly. This becomes a problem for the speculator when the good which she can readily exchange (say, shares) is one for which the reservoirs do not have disparate prices, or in which they do not trade at all.»*

According to this model, agents capable of storing small amounts of a conserved product interact alternatively with two systems having differing *temperatures* with respect to the conserved product's parameters. This alternating interaction gives rise to an engine capable of extracting a conserved product from the flow between the two systems mediated by an agent. The total *work* done is simply the area inscribed by the system's trajectory. By *work* we mean the quantity of product taken from the source system but not delivered to the sink system and thus diverted for some unspecified purpose.

Some important observations can be made about systems that behave in this manner. In particular

1. The process followed by such an engine is not perfectly efficient and introduces entropy during every cycle. As a result, such an engine is not fully reversible and the interactions between the two systems are therefore not fully reversible.

2. The definition of *work* is based on the product transfer process and not on properties of the agent itself. Thus the difference in any two states of the systems depends on the path taken and not on the states themselves. Unlike entropy and *temperature*, the *work* done is not an intrinsic property of systems. In an economic system, one can think of *work* as the source of profit.

3. The use of such a model requires us to define something like *free product*–that fraction of the total product available that may be ideally extracted from a system using such an engine. As with the notion of free energy, free product is a very important intrinsic property of any complex system.

4. The efficiency of such an engine is determined exclusively by the difference between the *temperatures* of the systems it is connecting to, and is not a property of the engine itself. Otherwise, one can show that non-conserving agents could be constructed, which is impossible or at best unsustainable.

5. Such an engine can be run in reverse and can use product to move product in the opposite direction it would naturally follow based on the *temperature* gradient. In an economic system, one can think of the product used as investment in the system of higher *temperature* and exploitation of the system with lower *temperature*.

The generalization of this model to military, law-enforcement, and engineering systems could be very helpful to understanding some of the most challenging phenomena we observe.

## BIOLOGY CHALLENGES

Entropy computations play an important role in understanding spatial patterns of gene expression, waves of low chemical concentrations in circadian clocks and along life-sustaining chemical pathways. Entropy also plays a critical role in understanding from a statistical perspective how noise in chemical concentrations and locations enhance and amplify the many weak signals that are essential to a healthy cell.

The sequencing of the human genome opens the door for vast numbers of new drug targets. It also increases the complexity of the task because it enables the study of drug interactions. Drug development is an expensive design-oriented process, and methods are needed to select effective drug targets as early as possible. Fuhrman *et al.* [30] argue that time series from gene expression data with a high entropy are more likely to be effective drug targets. Langmead *et al.* [31] also use entropy-based methods (including the Shannon extension in information processing) to detect periodic expression patterns in gene expression data. Nemenman *et al.* further extend this approach to address biological time-series problems:

*«The major problem in information theoretic analysis of neural responses and other biological data is the reliable estimation of entropy—like quantities from small samples. [A recently introduced Bayesian entropy estimator...] performs admirably even very deep in the undersampled regime, where other techniques fail. This opens new possibilities for the information theoretic analysis of experiments, and may be of general interest as an example of learning from limited data.»* [32]

Indeed, timing is critical in many biological processes, but it is measured differently at different scales, ranging from the frequency of chemical waves at the cellular level up to the variations of daylight and ambient temperatures at the organism level. At every level, the clocks must be synchronized. Time can only be measured at any level because some change occurred, e.g. in chemical concentrations or in energy flux. Biological macro-systems may therefore have a faster clock cycle than some of their subsystems. Rojdestvenski and Cottam [33] show how the concept of entropy can be used to estimate the difference of clock speed based on the number of states in the systems. Furthermore, according to Vilar *et al.* [34] noise can actually enhance the weak signals that drive the clocks:

*«A wide range of organisms use circadian clocks to keep internal sense of daily time and regulate their behavior accordingly. Most of these clocks use intracellular genetic networks based on positive and negative regulatory elements. The integration of these "circuits" at the cellular level imposes strong constraints on their functioning and design. [… This] type of oscillator is driven mainly by two elements: the concentration of a repressor protein and the dynamics of an activator protein forming an inactive complex with the repressor. Thus the clock does not need to rely on mRNA dynamics to oscillate, which makes it especially resistant to fluctuations. Oscillations can be present even when the time average of the number of mRNA molecules goes below one. Under some conditions, this oscillator is not only resistant to but paradoxically also enhanced by the intrinsic biochemical noise.»*

In fact, it has been known for a decade that the response of non-linear systems to weak signals may be enhanced by the presence of noise; such systems include animal feeding behavior, human tactile and visual perception and some neurobiological systems, see [35] and references. Such phenomena are examples of the so-called stochastic resonance statistical dynamics. Perc and Marhl [36] rigorously studied stochastic resonance effects in the response of intracellular $Ca^{2+}$ to weak signals. They show how different noise intensities enhance optimally different signal frequencies. Kummer and Ocone [37] show how the concept of entropy leads to a thermodynamic formulation of the epigenic system in which the pseudo (or talandic) temperature measures the chemical oscillations between genetic locus and the resulting metabolites. The model used is a simplified one but it suggests how entropic processes play a central role in understanding the adaptive behavior complex macro-scopic biological systems.

Another area where the interplay between system entropy and conservation laws may play a major role is cellular morphology, the study of how stem cells either proliferate or generate specialized descendants [38]. Silva and Martins [39] apply entropy related concepts within a cellular automata framework and use a model of gene coupling to study the frequency of specialization.

The above examples illustrate that entropic approaches to modeling complex biological systems may yield new insights and accurate predictions achievable neither by entirely atomistic models nor by macro-scale bulk models. Two challenges emerge also. First, biological systems are inherently hierarchical with coupling across levels. Entropy estimates at one level must be related to entropy estimates at the next level. One illustration of this need is that whereas the fast interactions between small molecules can be accurately modeled using reaction rates, the interactions with and between proteins involves relatively slower conformational changes and need not be accurately modeled using simple reaction rates [40]. Deterministic models of multi-scale systems face the same challenge; however, in the present case the coupling is empirical. Cross-design for very large statistical linear models with 100,000,000 equations is currently being investigated by Abowd [41]. Thus the first hurdle may not be so daunting.

A second challenge is that most biological processes appear discontinuous at a small enough time scale. The last 100 years have seen the development of mathematical analytical techniques that address specifically periodic phenomena; most biological processes in a healthy organism are cyclic. Drawing on the experience gained from mathematical physics, basic research is needed to build accurate quantitative models of realistic biological systems. As Patel *et al.* [42] put it in the context of a quantum DNA search problem:

*«Identification of a base-pairing with a quantum query gives a natural (and first ever) explanation of why living organisms have 4 nucleotide bases and 20 amino acids. It is amazing that these numbers arise as solutions to an optimisation problem.»*

## NATIONAL SECURITY CHALLENGE

Small worlds models of social networks have produced a great deal of literature on the subject of punctuated equilibrium dynamics. Bak and Sneppen [43] first brought together the concept of punctuated equilibria and self-organizing criticality in addressing complex evolutionary systems. The parallel between species in an ecological environment and clusters in a social network seems to suggest itself without much effort.

Recent work comparing the properties of social clusters to those of spin clusters [44] has shown parallels between phase changes in market and labor stratifications, and determined the requisites for redrawing the membership boundaries with respect to energy minimization and efficiency. In particular, this work has shed some light on quantifiable variables relating to the concept of "self" and its importance in the evolution of social clusters. Distorted ideas of self can account for altruism in such systems, and by extension one expects them to account for other forms of extreme individual behavior including sociopathy.

The infrequent upsurge of anomalous and extreme social behavior can also be approached by studying phenomena such as the onset of synchronization in networks of coupled oscillators [45]. Averaging theory has been used to determine the ranges of parameters for which network oscillations are damped as the system approaches a global steady state [46]. The application of this to any periodic process and in particular to social phenomena is readily apparent.

By simple extension of the Bak–Sneppen model, we can see that the stability of a social cluster is determined by the barrier height separating its local fitness maximum from other better maxima. In the case of an individual in a social network, the barrier height might simply be the number of beliefs that must be changed. Single belief changes occur often, but complicated changes such as radical multi-dimensional belief revisions are prohibitively unlikely because they involve complex and coordinated moves through an ill-defined landscape. Thus, the time scale for change is exponential in the barrier height. When the fitness of belief is high, it is often very difficult to find a nearby maximum that is better. Those states are relatively stable. When the fitness is low, it is very likely that a nearby better state can be found, so the barriers are low. The challenge for modelers of social networks in the context of sociopathic behavior is that the ability of individuals to quickly find erroneous maxima has been greatly increased by the widespread availability of dubious information and the increasing resonance of provocative messages.

Without much comment on its obvious historical impact, Carvalho-Rodriguez et al. [47] studied the loss of social cohesion resulting from pandemics between the 14$^{th}$ and 17$^{th}$ Centuries in Europe. Based on the data, they suggest that a critical threshold exists near 37% mortality, at which point the disruption of the social structure is maximal. But they also suggest using this entropy measure to gain insight into the impact of low-intensity conflict on the stability of civil government.

According to Herman [48] conflict models are fundamentally attrition-based, and entropy is the macroscopic expression of the combined effect of Clausewitzian friction, disruption, and lethality. Taken from another perspective, entropy collectively expresses unit cohesion and capability—as a unit loses cohesion its entropy rises and its capabilities decrease. Combat casualty models have also been developed based on entropic processes. Dexter [49] investigated the use of casualty based entropy and the entropic phase space developed by Carvalho-Rodrigues et al. [50], to determine their usefulness as measures of combat effectiveness. Dexter extended the combat entropy method (an entropy difference) for time-independent data to show its utility for time-dependent data, where Carvalho-Rodrigues applied an entropic phase space. While the entropic phase space was determined to be a poor outcome predictor, Dexter found that the combat entropy between two forces is a good predictor of attrition-based outcomes wherein the successful force is usually one that can maintain lower entropy for a greater time period.

Models such as this show a great deal of promise for detecting and monitoring anomalous social and organizational behavior. However, we are a long way from implementing these in practical systems. Furthermore, visualization of social network analysis systems presents an awesome challenge that has yet to be addressed satisfactorily. Reviewers of the subject often suggest that visualization is critical to the progress of scientific fields [51]. Freeman discusses at length how computing technology has affected the study of social systems, including procedural analysis methods that make use of the interaction of viewers with network data to explore their structural properties. This approach has taken root in the work done by the leading visualization research groups around the world, including those at research laboratories of the Department of Energy, National Institutes of Health, Department of Commerce, Department of Defense, and various law enforcement agencies. It is clear that visualization technology is an essential element to moving the existing theoretical body of knowledge about the behavior of complex adaptive systems into practical domain-specific and application-oriented technology suites that can address the most pressings problems of the day.

## ENGINEERING CHALLENGE

System complexity is a long-term challenge to the design and operation of many engineered systems, particularly those that interact with market-based financial systems, such as the transportation logistics network, telecommunications networks, and the electric and gas energy delivery networks. Developing tools and techniques to manage system complexity are essential to a) developing an understanding of how technologies, policies, and regulations will impact the stability of complex, evolutionary, economically driven systems; and b) monitoring markets in real-time to ensure they are not being manipulated by the "invisible hand" of a cheater who violates the rules.

The phenomena that must be studied are comprised of multiple independent devices and agents operating at a variety of scales – physical scales ranging from continents down to devices within individual homes and businesses, and time scales ranging from milliseconds to decades. For example, when the electric grid actively engages end-use appliances and equipment as integral elements of system control, the scale of the U.S. system expands from roughly 100 thousand electrical busses to more than 1 billion pieces of equipment, a 10-thousand-fold growth. Transient stability phenomena are driven by sub-second electromechanical and control processes, while planning and construction of power plants, transmission lines and distribution substations takes place over the course of years and their economic lifetimes extend over decades. The ability to understand and manage these phenomena involves answering a variety of fundamental scientific challenges.

Recent research by Oliveira *et al.* [52] has extended mathematical methods originally developed for exploring the complex combinatorial problem of ultra-large biological networks [53] to the problems posed by the power grid. Known as Petri networks, this method relies on a decision network that uses mathematical equations to express complex and highly dimensional relationships and can be used to construct models. Petri nets have been used to efficiently and comprehensively pre-compute the boundaries of network stability regions, and the inter-network influence relationships, reducing the computational complexity associated with this class of network-of-networks that are hyper-exponentially hard to solve to near linear time complexity. Most importantly they have shown the ability to identify system invariants that correspond to conservation laws in the traditional engineering models.

Additional research has been conducted into an underlying theoretical basis for the behavior of most market-governed large-scale engineering system. Development of this theory of transactive control can be used to predict and manage complex engineering system behavior [54]. Similar to the way statistical thermodynamics is used to describe large numbers of molecular particles, the theory of transactive elements applies statistical methods to analyze the overall functional properties of populations of machines as they obey conservation laws for energy trading in a transactive energy grid. This approach already has revealed profound misconceptions about the behavior of loads in the presence of utility demand response programs [55]. It also has been shown to predict distributions of electricity prices that match observations from some of today's wholesale markets [56]. It is hoped that such a theory will eventually provide the basis for knowing whether a market is operating in an abnormal condition (i.e., being manipulated).

Ultimately, we need modeling approaches that allow for evolutionary changes to the strategies of behavioral agents to develop the decision support tools for policy makers and regulators who are concerned with preserving public and private benefits. The traditional deterministic approach used by engineering simulation is woefully inadequate, hence the need to understand better the role of entropic equilibrium and its relationship to system conservation laws. Experimental economics has given us some insights into what these conservation laws might be and how individual decision processes contribution to the entropy of engineered econophysical systems. Vernon Smith, winner of 2003 Nobel Prize for economics, pioneered techniques of experimental economics that used people as actors in a gaming environment to explore the distinction between actual behavior and theoretically ideal behavior. HyungSeon *et al.* [57] reported being successful at developing insights into the strategies and differences between automated and human agents as they operate in power markets. The next step is to incorporate the exploratory behavior of humans into the engineering models of system behavior such that the economic motives of player-agents are more accurately reflected in the long-term dynamics of the system. This will enable the study market structures and reveal how economic and technical control strategies co-evolve over time so that appropriate rules and policies for regulating the national infrastructures can be comprehensively explored.

## CONCLUSIONS

We have reviewed the importance of improved and more approachable analytic methods to study the wide variety of complex adaptive systems we wish to consider. This has not been to downplay the importance of numerical tools in their own right. Indeed, we believe it is necessary to reestablish a balance between these two approaches rather than substitute the former for the latter. Numerical tools offer essential insights, and they are often very useful in quickly identifying invalid or weaknesses in hypotheses. But ultimately, it is the deeply revealing insights gained from analytic methods such as statistical and dynamical analyses that lead to the most transformative discoveries. Our ability to brandish hugely powerful numerical methods on vast computational platforms is a double-edged sword; on the one hand, it has given us unrivaled capabilities to simulate the rational deterministic behavior of large-scale systems, but that has come at the expense of somewhat atrophied *ab initio* and statistical analytic capabilities. Clearly for us to advance the science of complex adaptive systems, we must reinvigorate these latter capabilities and bring them to bear on the problems we seek to solve.

This work was supported by Pacific Northwest National Laboratory, which is operated by Battelle Memorial Institute for the U.S. Department of Energy under Contract DE-AC06-76RL01830.